\documentclass[%
reprint,
 amsmath,amssymb,
 pre,
]{revtex4-1}

\usepackage{graphicx}
\usepackage{dcolumn}
\usepackage{bm}
\renewcommand{\d}{{\rm d}}
\newcommand{\e}{{\rm e}}
\newcommand{\PD}[2]{\dfrac{\partial #1}{\partial #2}}
\newcommand{\FD}[2]{\frac{\d #1}{\d #2}}
\graphicspath{{Figures/}}

\usepackage{todonotes}

\DeclareSymbolFont{AMSb}{U}{msb}{m}{n}
\DeclareMathSymbol{\NSet}{\mathalpha}{AMSb}{"4E}
\DeclareMathSymbol{\ZSet}{\mathalpha}{AMSb}{"5A}
\DeclareMathSymbol{\RSet}{\mathalpha}{AMSb}{"52}
\DeclareMathSymbol{\CSet}{\mathalpha}{AMSb}{"43}
\DeclareMathOperator{\real}{Re}
\DeclareMathOperator{\imag}{Im}

\usepackage{natbib}
\begin{document}


\title{A next generation neural field model: The evolution of synchrony within patterns and waves}

\author{\'{A}ine Byrne}
 \email{aine.byrne@nyu.edu}
 \affiliation{Center for Neural Science, New York University, New York, NY 10003 USA.}
 \affiliation{
Centre for Mathematical Medicine and Biology, School of Mathematical Sciences, University of Nottingham, University Park, Nottingham, NG7 2RD, UK.
}%
 \author{Daniele Avitabile}
\affiliation{
Centre for Mathematical Medicine and Biology, School of Mathematical Sciences, University of Nottingham, University Park, Nottingham, NG7 2RD, UK
}%
\affiliation{
  Inria Sophia Antipolis Méditerran\'ee Research Centre, MathNeuro Team, 2004 route
  des Lucioles–Boîte Postale 93 06902 Sophia Antipolis, Cedex, France.
}%
\author{Stephen Coombes}
\affiliation{.
Centre for Mathematical Medicine and Biology, School of Mathematical Sciences, University of Nottingham, University Park, Nottingham, NG7 2RD, UK.
}%

\date{\today}

\begin{abstract}
Neural field models are commonly used to describe wave propagation and bump attractors at a tissue level in the brain. Although motivated by biology, these models are phenomenological in nature. They are built on the assumption that the neural tissue operates in a near synchronous regime, and hence, cannot account for changes in the underlying synchrony of patterns. It is customary to use spiking neural network models when examining within population synchronisation. Unfortunately, these high dimensional models are notoriously hard to obtain insight from. In this paper, we  consider a network of $\theta$-neurons, which has recently been shown to admit an exact mean-field description in the absence of a spatial component. We show that the inclusion of space and a realistic synapse model leads to a reduced model that has many of the features of a standard neural field model coupled to a further dynamical equation that describes the evolution of network synchrony. Both Turing instability analysis and numerical continuation software are used to explore the existence and stability of spatio-temporal patterns in the system. In particular, we show that this new model can support states above and beyond those seen in a standard neural field model. These states are typified by structures within bumps and waves showing the dynamic evolution of population synchrony.
\end{abstract}

\pacs{Valid PACS appear here}
\maketitle

\section{\label{sec:intro}Introduction}
The act of passing information between brain regions produces waves of neural activity. These waves are readily observed using non-invasive techniques such as electroencephalography (EEG) and magnetoencephalography (MEG) \citep{Alexander2015}, as well as in brain slices \citep{Golomb97}. Both experimental and theoretical work has shown that EEG/MEG recordings and evoked potentials can exhibit travelling and standing waves \cite{Nunez2006}. In particular travelling waves are seen in EEG sleep recordings propagating across the cortex at a speed of about 1.2-7.0 m/s \cite{Massimini2004}. Standing waves are often associated with idle brain states. For example, standing waves at $\alpha$ frequency (8-13 Hz) are observed in the vicinity of the visual cortex when the subject has their eyes closed \cite{Berger1929}. Another commonly observed spatial pattern is the so called \emph{bump attractor}. This spatially localised increase in population firing is produced in working memory tasks and the location of the bump can be linked to memory location \cite{Wimmer2014}.

Traditionally, neural field models are used to describe wave and bump states in the brain. Although inspired by biology, these models are entirely phenomenological in nature. Even so, they have been particularly successful in describing neurophysiological phenomena, such as EEG/MEG rhythms \citep{Zhang1996}, working memory \citep{Laing2002}, binocular rivalry \citep{Bressloff2012} and orientation tuning in the visual cortex \citep{Ben-Yishai1995}.
They are typically cast as a system of non-local differential equations which describe the spatiotemporal evolution of coarse grained population variables, such as the firing rate of a neuronal population, the average synaptic current, or the mean membrane potential \citep{Coombes2014}. 

The first attempt at a neural field mode is attributed to Beurle \cite{Beurle1956}, who built a model to describe the propagation of activation in a given volume of neural tissue. This model was purely excitatory, but even so allowed him to examine the propagation of large scale brain activity. In the 1970s Wilson and Cowan \cite{Wilson1972, Wilson1973} extended this model to include a second inhibitory layer. Unlike Beurle, they were interested in spatially localised bump solutions. In his seminal paper, Amari \cite{Amari1975,Amari1977} created what is now known as the standard neural field equation. By introducing a Mexican hat type coupling function (local excitation and long range inhibition), he reduced the model to a single equation with a mixture of excitatory and inhibitory connections. This allowed him to construct explicit solutions for spatially localised patterns, and assess their stability (at least for a Heaviside firing rate function).  For a review of the Amari model and \emph{bumps} in one spatial dimension see \cite{Coombes2005} and for a discussion of \emph{bubbles} in two spatial dimensions see \cite{Bressloff2013}.

One of the main assumptions in many types of neural field model, and especially those for describing EEG, is that point-wise they describe a density of neurons operating in a near synchronous regime \cite{Nunez2005}.  This wholly reasonable assumption can be traced back to the observation that an EEG scalp electrode, which typically experiences the activity of roughly $10^9$ cortical pyramidal cells, can only detect an electric field if all the individual cell dipoles add coherently \cite{daSilva05}.  However, a near synchrony assumption in any neural mass model precludes its use in describing the increase and decrease of power commonly seen in given EEG/MEG frequency bands.  These \textit{temporal} variations are believed to be the result of changes in synchrony within the neural tissue. The former phenomenon is called event-related synchronisation (ERS), and the latter event-related desynchronisation (ERD) \cite{Pfurtscheller1999}.  Consequently, there is a pressing need to develop the next generation of neural field models, which include this notion of a dynamic within-population synchrony (not fixed to be near synchronous), to more accurately describe the evolution of large scale spatio-temporal brain rhythms. 

When looking at within population synchronisation one typically uses a spiking neural network model. However, these high dimensional models are almost impossible to gain insight from.
In an ideal world there would be a mathematical procedure for linking microscopic dynamics to macroscopic dynamics. This link has proved elusive for the majority of spiking models. However, Luke \emph{et al.} \citep{Luke2013} showed that the $\theta$-neuron model is amenable to such a reduction for pulsatile coupling. Montbri\'o \emph{et al.} \citep{Montbrio2015} have used a similar approach to reduce a network of quadratic integrate-and-fire neurons. Laing \cite{Laing2015} has also shown that the same approach can be applied to a network of spatially extended $\theta$-neurons, in the presence of gap junction coupling. In previous papers  \citep{Byrne2017, Coombes2017}, we have shown that the approach of Luke \emph{et al.} can be extended to incorporate a biologically realistic form of synaptic coupling.
Here we build on this work to construct a neural field model that incorporates within population synchrony and a realistic form of synaptic coupling. We shall refer to this model as a \emph{next generation neural field model}. In this paper we show, using a mixture of analysis and simulation, that this new neural field model can support exotic patterned states more reminiscent of high dimensional spiking networks, with spatio-temporal patterns showing the evolution of synchrony.

We begin with an overview of the model formulation in \S\ref{sec:model} and outline the necessary steps for reduction to a neural field. A Turing instability analysis of the model is covered in \S\ref{sec:turing}. 
Here, we show that the system can be unstable to both static and dynamic Turing patterns for a wide window of parameter space. More interestingly, we show that when the Turing bifurcation collides with a Hopf bifurcation patterned states emerge in which there exists an oscillating structure within a spatially localised bump.
The Turing analysis is complimented with a numerical bifurcation analysis for the
full nonlinear model in \S\ref{sec:numeric}. Here we further examine the emergent
patterns away from bifurcation, as well as consider localised travelling waves.
Finally, in \S\ref{sec:discussion} we discuss our main results as well as natural extensions of the work presented.

\section{The Model} \label{sec:model}
\noindent We first consider a network of $N$ coupled quadratic integrate-and-fire (QIF) neurons, uniformly
distributed along a line of length $\Lambda$ such that the $j$th neuron is at position $x_j
= -\Lambda/2 +(j-1)\Delta x$, where $j=1,\ldots,N$ and $\Delta x = \Lambda/(N-!)$ is the spacing between neurons.
The coupling between neuron $i$ and neuron $j$ depends only upon the distance between the two
neurons, $w_{ij}^m=w_m(|x_i-x_j|)$, where $m$ is a label used to keep track of neural subpopulations. We shall focus on the choice $\Lambda \gg 1$ (so that we deal with large spatial scale systems).
We write the network dynamics for the voltage, $v_i \equiv v(x_i,t)$, in the form 
\begin{equation}
\dot{v}_i = v_i^2 +\eta_i + I_i , \qquad i=1,\ldots, N,
\end{equation}
subject to \textit{reset}, $v_i \rightarrow v_r$, whenever a firing threshold $v_{th}$ is reached by $v_i$.  The time at which the $i$th cell reaches threshold from below for the $s$th time will be denoted by $T_i^s$, $s\in \NSet$.
Here the background drives $\eta_i$ will be assumed to be heterogeneous and chosen from a Lorentzian distribution,
\begin{align}
L(\eta) = \frac{1}{\pi}\frac{\Delta}{(\eta-\eta_0)^2+\Delta^2},
\end{align}
where $\eta_0$ is the centre of the distribution and $\Delta$ is the half width.
We assume a synaptic input current of the form
\begin{equation}
I_i(t) = \sum_m g^i_m (t)(v_\text{syn}^m-v_i(t)),
\end{equation}
for a global conductance $g^i_m $, synaptic reversal potential $v_\text{syn}^m$ and local voltage $v_i$. 
In what follow we will assume $m=\{1,2\}$ or $m = 1$, but one should note that this framework can be extended to include multiple types of synapses.
The interplay of excitation and inhibition, on different spatial scales, is known to play an important role in the generation of global spatially patterned states \citep{Ermentrout1979,Bressloff2001,Coombes2010}.
Hence, we separate our synaptic conductance into two parts, such that we have  one excitatory synaptic current ($v_\text{syn}>0$) and one inhibitory ($v_\text{syn}<0$), each with different spatial ranges.

Each of the synaptic conductances will be taken to mimic that of a synapse with a finite rise and fall time that evolves according to
\begin{equation}
Q_m g^i_m  = \frac{\kappa_m}{N}  \sum_{s\in\NSet} \sum_{j=1}^N w_{ij}^m \delta(t-T_j^s),
\end{equation}
for some coupling strength $\kappa_m$, where $Q$ is the linear 2\textsuperscript{nd} order differential operator
\begin{equation}
Q_m = \left(1 + \tau_m\FD{}{t}\right)^2.
\label{eq:Q}
\end{equation}
Here $\tau_m$ is the synaptic time scale, and $Q_m$ has a response (Green's function) $s(t)=\tau_m^{-2} t \e^{-t/\tau_m}$ for $t\geq 0$ (and is zero otherwise), which is a popular choice for many synapse models in computational neuroscience \citep{Gabbiani2010}.

It is well known that the QIF model is formally equivalent to the $\theta$-neuron model \cite{Ermentrout1986} under the transformation $v_i = \tan(\theta_i/2)$, for $\theta_i \in [0, 2 \pi)$ (when the threshold $v_{th}$ and reset $v_r$ are set to $+\infty$ and $-\infty$, respectively). This relationship allows us to construct a $\theta$-neuron network dynamics as
\begin{align}
\dot{\theta_i} &= 1-\cos\theta_i + (1+\cos\theta_i)\eta_i \nonumber\\
&\phantom{.}\hspace{1.7em}+ \sum_m g^i_m  [(1+\cos\theta_i)v_\text{syn}^m - \sin\theta_i], \label{eq:theta_discrete} \\
Q_m g^i_m &= \frac{2 \kappa_m}{N}\sum_{j=1}^N w_{ij}^m\delta(\theta_j-\pi). \label{eq:g_discrete}
\end{align}
To obtain \eqref{eq:g_discrete} we have made use of the fact that $\delta(t-T_j^s) = \delta(\theta_j- \pi) |\dot{\theta}_j(T_j^m)|$.  As such, we say that neuron $j$ ``spikes" whenever $\theta_j$ increases through $\pi$.
The network formulation in terms of dynamics on a circle is particularly useful since we no longer have to worry about handling the discontinuous reset process as we would have to do for a QIF network.

\subsection{Mean field limit}
We take the large $N$ limit, $N \rightarrow \infty$, which allows us to describe the system in terms of a continuous probability distribution function $\rho(x,\eta,\theta,t)$, with $x,\eta \in \RSet$, $\theta \in [0,2 \pi)$, and $t\in \RSet^+$, which satisfies the continuity equation:
\begin{equation}
\PD{}{t}\rho +\PD{}{\theta}(\rho v_{\theta}) = 0,
\label{eq:continuity}
\end{equation}
where $v_{\theta}$ is the following realisation of \eqref{eq:theta_discrete},
\begin{align}
v_{\theta}= 1-\cos\theta + (&1+\cos\theta)\eta \nonumber \\
&+ \sum_m g_m [(1+\cos\theta)v_\text{syn}^m - \sin\theta] . \label{eq:theta_continuous}
\end{align}
The mean field representation of the synaptic inputs $g_m$ are written as follows,
\begin{align}
Q_m g_m (x,t) = \frac{\kappa_m}{\pi}\sum_{l \in \ZSet}\int_{-\infty}^\infty\d y & \int_0^{2\pi}\d\theta \int_{-\infty}^\infty\d\eta \rho(y,\eta,\theta,t) \nonumber \\
&\times w_m(x-y)\e^{il(\theta-\pi)}, 
\label{eq:g_continuous}
\end{align}
where we have used the result $2 \pi\delta(\theta-\pi) = \sum_{l \in \ZSet} \e^{il (\theta-\pi)}$ to write the right hand side in terms of exponentials. 
The formula for $v_{\theta}$ \eqref{eq:theta_continuous} may be conveniently written in terms of $\e^{\pm i\theta}$ as
\begin{equation}
v_{\theta} =  \phi (\e^{i\theta} +\e^{-i\theta}) + \chi,
\label{eq:c_exponential_form}
\end{equation}
where $\phi = (\eta+I-1)/2$ and $\chi  = \eta+I+1$.
Note that a similar approach has previously been considered by Laing \citep{Laing2015}. However, his focus was on smooth (non-pulsatile) interactions with a first order model of the synapse, and he did not consider reversal potentials.\\

To reduce the system we make use of the Ott-Antonsen (OA) ansatz \citep{Ott2008}.  This decomposes $\rho$ in the form
$\rho(x,\eta,\theta,t) = L(\eta) F(x,\eta,\theta,t)/(2 \pi)$, where $F$ is $2 \pi$-periodic in $\theta$ with a Fourier series representation $F(x,\eta,\theta,t) = \sum_n F_n(x,\eta,t)  \e^{in \theta}$.  The OA ansatz restricts the choice of $F_n$ such that 
$F_n(x,\eta,t) = \alpha(x,\eta,t)^n$, with $|\alpha(x,\eta,t)|<1$ to ensure convergence. Hence, $\rho$ can be written as
\begin{equation}
\rho(x,\eta,\theta,t) = \frac{L(\eta)}{2 \pi} \left\{ 1 + \left[\sum_{n=1}^\infty \alpha(x,\eta,t)^n \e^{i n \theta} + \text{cc} \right] \right\},
\label{eq:rho_OA}
\end{equation}
where $\text{cc}$ denotes the complex conjugate. Substituting \eqref{eq:c_exponential_form} into the continuity equation \eqref{eq:continuity} and balancing terms in $\e^{i\theta}$ gives an evolution equation for $\alpha$:
\begin{equation}
\PD{}{t} \alpha -  i \alpha^{2} \phi - i\alpha \chi - i \phi = 0 .
\label{eq:cont_a}
\end{equation}
We define the Kuramoto order parameter as follows,
\begin{equation}
z(x,t) = \int_0^{2 \pi} \d \theta \int_{-\infty}^\infty \d \eta  \rho(x,\eta,\theta,t)  \e^{i \theta},
\label{eq:kuramoto_order1} 
\end{equation}
where $|z|\leq 1$. The Kuramoto order parameter is a complex number $z=R\e^{i\Psi}$, whose magnitude $R$ represents the degree of within population synchrony and angle $\Psi$ represents the average phase of the population.
Substituting \eqref{eq:rho_OA} into \eqref{eq:kuramoto_order1} we find that
\begin{equation}
\bar{z}(x,t) = \int_{-\infty}^\infty \d \eta  L(\eta) \alpha(x,\eta,t),
\label{eq:kuramoto_orderr2}
\end{equation}
where $\bar{z}$ denotes the complex conjugate of $z$.
As the Lorentzian has two simple poles $\eta_\pm=\eta_0\pm i\Delta$, the above integral may be performed by choosing a large semi-circular contour in the lower half $\eta$-plane and using the residue theorem, to yield $\bar{z}(x,t) = \alpha(x,\eta_-,t)$.\\

\begin{figure}
\centering
\includegraphics[width=0.9\linewidth]{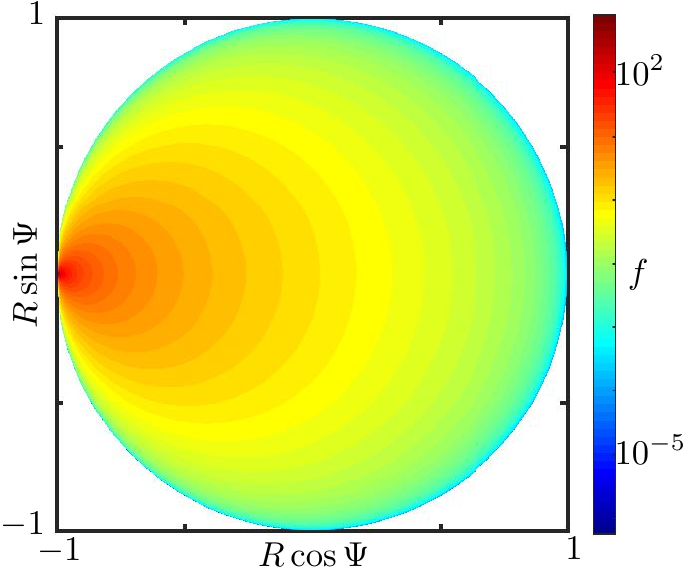}
\caption{\textbf{Firing rate dynamics:} Density plot showing the firing rate $f$ as a function of the complex Kuramoto order parameter $z=R\e^{i\Psi}$. Firing is highest near $z=\e^{i\pi}$. This corresponds to highly synchronous behaviour where all of the phases of the neurons go through $\pi$ simultaneously.
\label{fig:firing_rate}}
\end{figure}

We use \eqref{eq:kuramoto_order1} to write \eqref{eq:g_continuous} as 
\begin{equation} 
\begin{aligned}
Q_m g_m (x,t) &= \kappa_m\int_{-\infty}^{\infty} \d y  w_m(x-y) f(z(y,t)),\\
& \equiv \kappa_m w_m\otimes f(z),
\label{eq:evolution_g}
\end{aligned}
\end{equation}
where $\otimes$ represents a spatial convolution and $f$ is the population firing rate
\begin{equation} 
\begin{aligned}
f(z) &=  \frac{1}{\pi}\left\{1+\left[\sum_{l=1}^\infty (-1)^l z^l + \text{cc} \right]\right\}  \nonumber\\
&=  \frac{1}{\pi}\frac{1-\left| z\right|^2}{1+z+\overline{z}+\left|z\right|^2}, \hspace{1em} |z|<1.
\label{eq:firing_rate}
\end{aligned}
\end{equation}
Note that \eqref{eq:evolution_g} takes the form of a generalised neural field
equation, where the firing rate function $f$ is a derived quantity that depends on
the within population synchrony. Also noteworthy is the fact that this firing rate
function is not a sigmoid. It is a highly non-linear function that depends on the
intrinsic population dynamics. The
firing rate can be plotted as a function of the Kuramoto order parameter (Fig.~\ref{fig:firing_rate}). As expected, the firing rate is highest at
$z=\e^{i\pi}$ (where a singe neuron fires as $\theta$ increases through $\pi$).

The dynamics of $z$ are obtained by evaluating \eqref{eq:cont_a} at $\eta_- = \eta_0 - i\Delta$, and taking the complex conjugate, which gives $\partial z /\partial t = \mathcal{F}(z;\eta_0,\Delta)+\sum_m\mathcal{G}(z,g_m;v_{\text{syn}}^m)$, where 
\begin{align}
\mathcal{F}(z;\eta_0,\Delta)&=   -i\frac{(z-1)^2}{2}+\frac{(z+1)^2}{2}\left[i \eta_0 -\Delta \right],
\label{eq:F_internal} \\
\mathcal{G}(z,g;v_{\text{syn}})&=  g \left[i\frac{(z+1)^2}{2}v_{\text{syn}} -  \frac{(z^2-1)}{2} \right] .
\label{eq:G_external}
\end{align}
As in \cite{Coombes2017}, we interpret (\ref{eq:F_internal}) as describing the intrinsic population dynamics and (\ref{eq:G_external}) as the dynamics generated by synaptic coupling.

In summary, we have the following evolution equations
\begin{align}
&\partial_t  z = \mathcal{F}(z) + \sum_m\mathcal{G}_m(z,g_m) ,
\label{eq:ZEvol}
\\
&(1+\tau_m\partial_t)^2  g_m = \kappa_m w_m \otimes f(z),
\label{eq:GEvol}
\end{align}
where $\mathcal{G}_m(z,g_m)\equiv \mathcal{G}(z,g_m;v_{\text{syn}}^m)$ and we have omitted the dependence on control parameters. The dynamical
system~\eqref{eq:ZEvol}--\eqref{eq:GEvol} has therefore one complex ($z$) and $|m|$
real ($g_m$) state variables.
In what follows, we choose normalised exponentially-decaying synaptic kernels of the form $w_m(x) = \exp(-\beta_m |x|)/(2\beta_m)$, with Fourier transform $\widehat w_m(k) =1/(1+(k/\beta_m)^2)$. As the Fourier transform is a rational function, the evolution
equations~\eqref{eq:ZEvol}--\eqref{eq:GEvol} are equivalent to the PDEs \citep{Coombes2003}
\begin{align}
&\partial_t  z  = \mathcal{F}(z) + \sum_m \mathcal{G}_m(z,g_m) ,
\label{eq:ZPDE}
\\
&(1-\partial_{xx}/\beta_m^2) (1+\tau_m\partial_t)^2  g_m = \kappa_m f(z).
\label{eq:GPDE}
\end{align}

In what follows, we will show that the model can support Turing patterns, travelling waves, and other complex global spatio-temporal patterns, for $m=\{1,2\}$ and travelling fronts for $m=1$.
In the former case, this is achieved by choosing $\beta_1>\beta_2$ such that $w_1$ has a shorter spatial scale than $w_2$. For $v_\text{syn}^1>0$ and $v_\text{syn}^2<0$ this has the overall effect of short range excitation and long range inhibition, and for $v_\text{syn}^1<0$ and $v_\text{syn}^2>0$ inhibition dominates at short distances and excitation at longer distances.

\section{ Instability of the homogeneous steady state} \label{sec:turing}
We require excitation and inhibition to generate Turing patterns, hence we let $m=\{1,2\}$ in this section. We will also assume $\beta_1=1$ (without loss of generality) and define $\beta=\beta_2<1$ as the parameter which measures the difference in the spatial scales of $w_1$ and $w_2$.
We first recast the system \eqref{eq:ZPDE}--\eqref{eq:GPDE} as a first-order evolution equation in the
state variables $u = (a,b, K_1, g_1, K_2, g_2)$, where $a = {\rm Re}(z)$, $b = {\rm Im}(z)$ and $K_m = (1+\tau_m \partial /\partial
t)g_m$, and seek stationary and spatially homogeneous states $u(x,t) = u^*$
for all $x$ and $t$.

We apply a small perturbation of the form $\widetilde u(x,t) = u^* + \widetilde u(x,t)$, where $
\widetilde u(x,t) = A \e^{\lambda t} \e^{ikx}$, $\lambda \in \mathbb{C}$, $k\in \mathbb{R}$ and $A\in  \mathbb{C}^6$. Using the
following identity
\[
 w_m(x)\otimes\e^{ikx} = \widehat{w}_m(k)\e^{ikx} ,
\]
%
we obtain, to leading order,
$
\partial_t \widetilde u(x,t)  = \mathcal{J}(k)\widetilde u(x,t),
$
where $\mathcal{J}$ is the following ($k$-dependent) Jacobian, evaluated at $u^*$,
\begin{equation*}
\mathcal{J}(k) = \left(\begin{matrix}
\mathcal{J}_{11} & \mathcal{J}_{12} \\
\mathcal{J}_{21}(k) & \mathcal{J}_{22}
\end{matrix}\right),
\label{jacobian1}
\end{equation*}
where we highlight the $k$-dependence
\begin{equation*}
\mathcal{J}_{21}(k)  = \left(\begin{matrix}
\tau_1^{-1}\kappa_1 \widehat{w}_1(k)\partial_a  f \hspace{0.5em} \phantom{.}& \tau_1^{-1}\kappa_1 \widehat{w}_1(k)\partial_b  f \vspace{0.5em}\\
0 & 0 \vspace{0.5em}\\
\tau_2^{-1}\kappa_2 \widehat{w}_2(k)\partial_a f  \hspace{0.5em} \phantom{.}& \tau_2^{-1}\kappa_2 \widehat{w}_2(k)\partial_b f\vspace{0.5em}\\
0 & 0
\end{matrix}\right).
\label{jacobian2}
\end{equation*}
and show the other elements in Appendix \ref{app:jacobian}.

The complex eigenvalues $\lambda = \nu + i\omega$ satisfies the characteristic equation
\begin{equation}
\mathcal{E}(\lambda,k) = \det{\left| \mathcal{J}(k)-\lambda I_4\right|} = 0,
\label{eq:characteristic_formula}
\end{equation}
where $I_4$ is the $4\times4$ identity matrix. A homogeneous steady state $u^*$ is
linearly stable to perturbations $\e^{ikx}$ if $\nu(k) < 0$ for all
$k$. By the implicit function theorem, a branch of solutions $\lambda(k)$
to~\eqref{eq:characteristic_formula} touches the imaginary axis when
\begin{equation}
\partial_k \mathcal{M} \partial_\omega \mathcal{N} - \partial_\omega \mathcal{M} \partial_k \mathcal{N} = 0,
\label{eq:nonDegenracy}
\end{equation}
where $\mathcal{M}={\rm Re}(\mathcal{E})$ and $\mathcal{N}={\rm Im}(\mathcal{E})$.

\begin{figure*}
\centering
\includegraphics{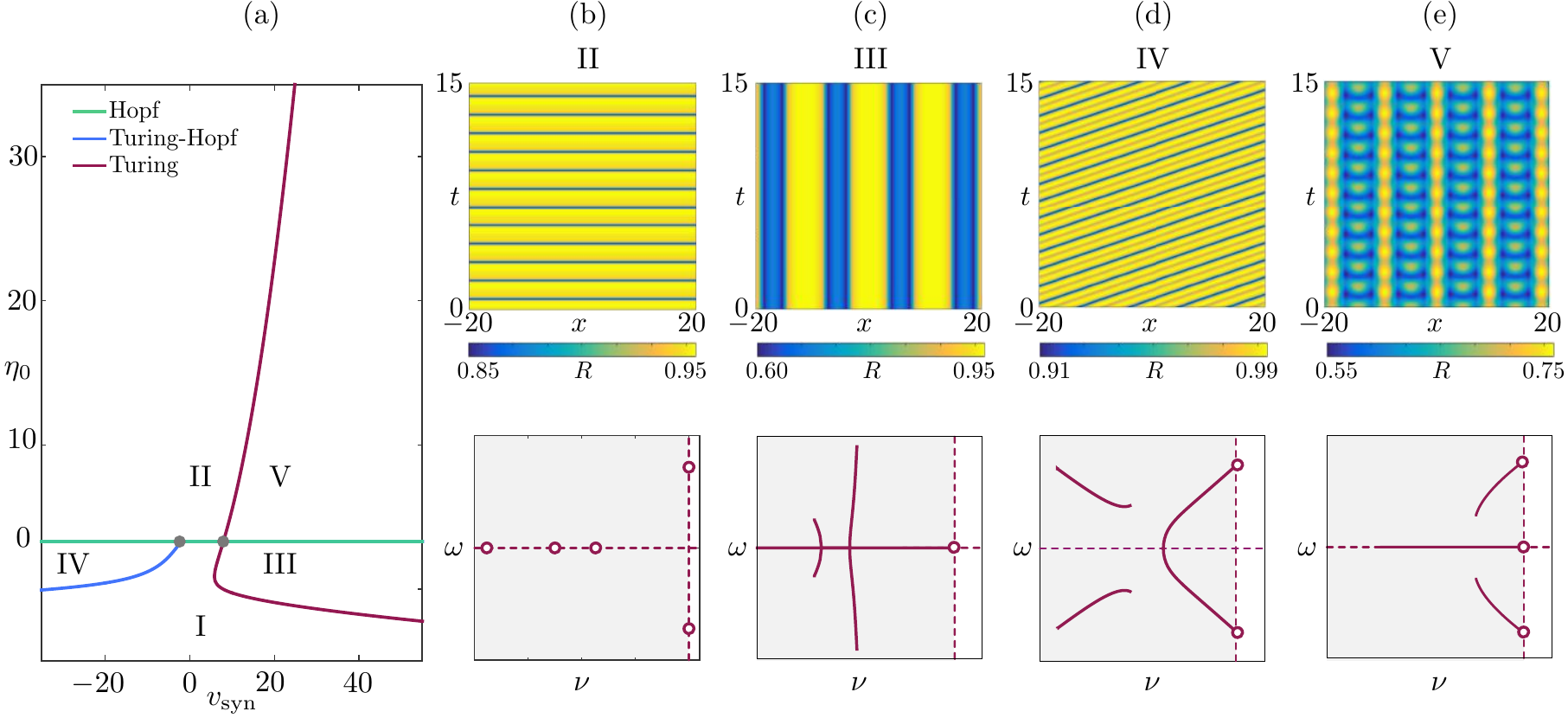}
\caption{Instabilities of the homogeneous steady state. (a): Two-parameter
bifurcation diagram of Hopf and Turing instabilities of the homogeneous steady state in the
$(v_{\text{syn}},\eta_0)$-plane, where $v_{\text{syn}}=v_{\text{syn}}^1=-v_{\text{syn}}^2$. The Hopf (green), Turing (red) and Turing-Hopf (blue) curves partition
the plane into five sectors. The Hopf and Turing curves cross at a codimension-2 point where 
both instabilities occur simultaneously. The homogeneous steady state 
$u^*$ is stable in I. (b): $u^*$ is unstable to global periodic oscillations in II, 
and direct numerical simulations close to the instability show a stationary Turing pattern. 
We plot $R(x,t) = |z(x,t)|$ obtained via direct numerical simulation 
of~\eqref{eq:ZPDE}--\eqref{eq:GPDE}, for $m=\{1,2\}$ with exponentially-decaying kernels 
(top panel). The eigenvalues of the spatially clamped system at the bifurcation point are 
shown in the bottom panel. (c): $u^*$ is Turing unstable in III, hence, we observe stationary 
periodic patterns. The spectrum of the linearised operator around $u^*$ at bifurcation
is also shown (bottom panel). (d): $u^*$ is Turing-Hopf unstable in IV and the
instability (bottom panel), gives rise to a periodic wavetrain, as expected. (e):
spatio-temporal pattern obtained in V, where $u^*$ is Turing and Turing-Hopf
unstable. The spectrum in the bottom panel is at the codimension-2 point, marked in
(a) with a red circle. Parameters: $\Delta=0.5$, $\kappa_1=\kappa_2=5$, $\tau_1=\tau_2=0.2$. }
\label{fig:turing_curves}
\end{figure*}

We observe a Hopf bifurcation of the spatially uniform state $u^*$ if \eqref{eq:characteristic_formula} holds for $\nu(0)=0$ and $k=0$, i.e. if there is a non-zero solution of:
\begin{align}
  \omega^6 + p_4\omega^4 + p_2 \omega^2 + p_0 &= 0, \label{eq:Hopf1}\\
  p_5 \omega^4 + p_3 \omega^2 +  p_1 &= 0. \label{eq:Hopf2}
\end{align}
Here, $p_i$ are scalars which depend on $u_*$ and the control parameters of the problem (see Appendix \ref{app:coeff}).
Solving \eqref{eq:Hopf1}--\eqref{eq:Hopf2}, for fixed width of the Lorentzian $\Delta$, synaptic
coupling strengths $\kappa_1$, $\kappa_2$, synaptic time constants $\tau_1$, $\tau_2$, 
and relative width of the synaptic kernel $\beta$, gives the locus
of Hopf bifurcations as a function of the synaptic reversal potential $v_\text{syn}$ and 
background drive $\eta_0$ (Fig \ref{fig:turing_curves}(a) green curve). We fix the synaptic reversal potentials to be equal and opposite, and define $v_\text{syn} = 
v_\text{syn}^1 = -v_\text{syn}^2$. For simplicity we set $\kappa_1=\kappa_2$ and 
$\tau_1=\tau_2$. For this choice of parameters, excitation dominates for short range 
interactions for $v_\text{syn}>0$ and inhibition dominates for short range interactions 
for  $v_\text{syn}<0$. Note, also, that the Hopf bifurcation occurs for the same value of $\eta_0$ for all 
$v_\text{syn}$. This is a consequence of the choice of equal coupling strengths and time constants. If this balance is disrupted the Hopf bifurcation will depend upon 
$v_\text{syn}$ and for some parameter choices we see several Hopf curves in the $(v_{\text{syn}},\eta_0)$-plane.

The homogeneous steady state $u^*$ undergoes a static Turing bifurcation if
there exists a non-zero critical wavenumber $k_c$ such that \eqref{eq:characteristic_formula} and
\eqref{eq:nonDegenracy} hold for $\nu(k_c)
=0$, $\omega(k_c)=0$. This leads to the conditions
\begin{align}
   &p_0 + q^1_0 \widehat w_1  + q^2_0 \widehat w_1    = 0, \label{eq:Turing1}\\
   &\left[q_0^1\FD{\widehat w_1}{k_c} +q_0^2\FD{\widehat w_2}{k_c}\right]\left(p_1+q^1_1 \widehat w_1  + q^2_1 \widehat w_2 \right)= 0, \label{eq:Turing2}
\end{align}
where $\widehat{w}_i$ and its derivative depend on $k_c$ and $\beta$, and
the scalars $p_i$, $q_i^j$ depend on $u_*$ and the control parameters of the
problem (see Appendix \ref{app:coeff}). 

As with the Hopf bifurcation, the locus of Turing bifurcations can be plotted as a function 
of the synaptic reversal potential $v_\text{syn}$ and background drive $\eta_0$ (Fig.~\ref{fig:turing_curves} red curve). The curve exhibits a turning point, 
therefore the system has two Turing bifurcations for sufficiently large values of
$v_\text{syn}$. If we fix $v_\text{syn}$ at a value in this region, and ascend the bifurcation diagram by increasing $\eta_0$, the spectrum
$\lambda(k)$ touches the imaginary axis of the $(\nu,\omega)$-plane with wave number $k_c$ (lower Turing bifurcation) (Fig. \ref{fig:turing_curves}(c)) and continues to move to the right, implying that $u^*$ is
unstable to a range of perturbations with wavenumbers $k \in (k_1,k_2)$. As $\eta_0$
is further increased, the spectrum returns to the left-hand plane (upper Turing
bifurcation), which restores the stability of $u^*$. Note the value of $k_c$ is not necessarily the same on the upper and lower branches.
The scenario described above is robust to perturbations in other
parameters: we found that changes in $\tau_1$, $\tau_1$ and $\Delta$ do not
significantly affect the location of the Turing bifurcations. Increasing $\kappa_2$
results in an increase of the gap between the two critical values of $\eta_0$, and
hence a larger window of instability, while increasing $\kappa_1$ decreases this gap and at a certain point the shape is inverted, such that the unstable region lies predominantly in the $\eta_0<0$ region of the plot. As $\kappa_1$ corresponds to the synaptic strength for the excitatory current for $v_\text{syn}>0$, when it becomes significantly larger than $\kappa_2$ excitation dominants even for long range interactions. As patterns arise from the interplay of excitation and inhibition, an inhibitory external drive ($\eta_0<0$) is needed to observe Turing patterns in this regime.

A Turing-Hopf instability of the homogeneous steady state occurs if there exists a non-zero wave
number $k_c$ such that \eqref{eq:characteristic_formula} and
\eqref{eq:nonDegenracy} hold with $\nu(k_c)=0$ and $\omega(k_c) = \pm \omega_c \neq
0$. This instability, which we will also refer to as \textit{dynamic} Turing
bifurcation, elicits wavetrains with wavenumber $k_c$ and phase velocity $\omega_c/k_c$ (near bifurcation).

From \eqref{eq:characteristic_formula} and \eqref{eq:nonDegenracy} we
obtain a system of the form
\begin{align}
& \omega^6 +p_4 \omega^4 + P_2\omega^2 + P_0 = 0, \label{eq:dynTuring1}\\
      &  p_5\omega^4 +    P_3 \omega^2 + P_1 = 0, \label{eq:dynTuring2}\\
&  \left[ Q_2^1\FD{\widehat w_1}{k} +Q_2^2\FD{\widehat w_2}{k}\right]                (-6\omega^5 +4 p_4 \omega^3 - 2 P_2\omega) \nonumber \\
 &\phantom{.}\hspace{0.5em}-\left[ Q_3^1\FD{\widehat w_1}{k} +Q_3^2\FD{\widehat w_2}{k}\right](5p_5\omega^4 -3 P_3 \omega^2 + P_1)  = 0, \label{eq:dynTuring3} 
\end{align}
where $P_i = p_i +q^1_i \widehat w_1(k)  + q^2_i \widehat w_1(k) $, $Q_i^j =q_i^j\omega^i +q_{i-2}^j\omega^{i-2}$ and $p_i$, $q_i^j$ are scalars which depend on the control parameters (see Appendix \ref{app:coeff}).
In passing, we note that the characteristic equation now has an imaginary
part, resulting in the two conditions~\eqref{eq:dynTuring1}, \eqref{eq:dynTuring2}.
Solving \eqref{eq:dynTuring1}--\eqref{eq:dynTuring3} allows us to plot the locus of the Turing-Hopf
bifurcations in the $(v_{\text{syn}},\eta_0)$-plane (Fig.~\ref{fig:turing_curves} blue curve), together with the Hopf and Turing bifurcation curves. 
Note that the Turing-Hopf curve intersects with the Hopf curve at $v_{\text{syn}}\approx 8$. The value of $k_c$ deceases along the Turing-Hopf curve as $v_{\text{syn}}$ is increased, such that at the point where the two curves collide $k_c=0$ on the Turing-Hopf curve.

\begin{figure*}
\centering
\includegraphics{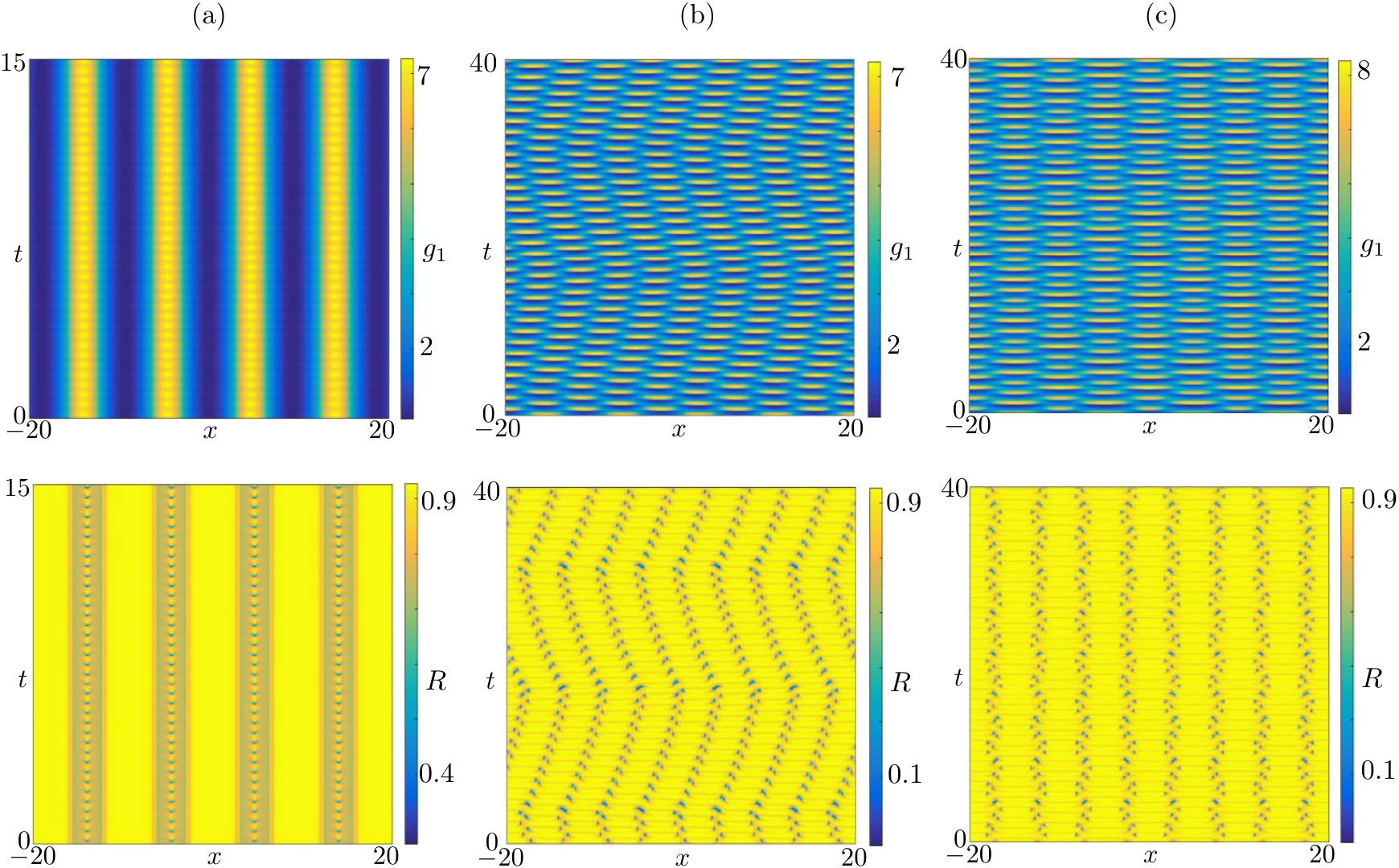}
\caption{Further spatio-temporal patterns supported by the neural field model. (a): 
  Pattern found in sector V of Fig.~\ref{fig:turing_curves}(a), for $\eta_0=60$,
  $v_\text{syn}^1=-v_\text{syn}^2=10$. The pattern displays a time-periodic structure within each 
  bump of a Turing pattern. (b): \emph{Wandering} bump observed in region IV of 
  Fig.~\ref{fig:turing_curves}, for $\eta_0=2.5$, $v_\text{syn}^1=-v_\text{syn}^2=-20$. 
  Similar patterns are also found in sector II (not shown). (c): Weakly unstable standing wave 
  obtained for $\eta_0=15$, $v_\text{syn}=-10$.
  After undergoing a breathing instability for a long transient, the standing wave evolves
  towards a stable wavetrain (not shown). Other parameter values as in Fig \ref{fig:turing_curves}.}
\label{fig:patten_zoo}
\end{figure*}

The curves in Fig.~\ref{fig:turing_curves}(a) were computed using {\sc
XPPAUT} \cite{Ermentrout02}, by continuing a suitable
algebraic problem in one of the parameters of the system ($v_\text{syn}$). The equations defining an equilibrium are solved simultaneously
with \eqref{eq:Hopf1}--\eqref{eq:Hopf2}, \eqref{eq:Turing1}--\eqref{eq:Turing2} or
\eqref{eq:dynTuring1}--\eqref{eq:dynTuring3} and continued in parameter space.
The curves partition the $(v_{\text{syn}},\eta_0)$-space into five sectors (labelled I--V). 
In addition to sectors where $u^*$ is stable (I), $u^*$ is unstable to bulk oscillations (II), Turing instabilities (III), and Turing-Hopf instabilities (IV), a fifth sector (V) is generated by the crossing of the Turing and
Hopf curves. At the intersection (codimension-2) point, the spectrum of the
linearised operator around $u^*$ has one zero eigenvalue and two complex conjugate
eigenvalues, where the critical wavenumber $k_c$ is non zero for the zero eigenvalue and equal to zero for the complex conjugate pair (Fig \ref{fig:turing_curves}(e)). 
Direct numerical simulations
close to the instability confirm the predictions of the linear stability analysis:
bulk oscillations are observed in in section I (Fig \ref{fig:turing_curves}(b)) 
stationary Turing patterns are observed in sector III (Fig \ref{fig:turing_curves}(c)) and both wavetrains and standing waves are seen in region IV (Fig \ref{fig:turing_curves}(d)). 
In sector V, the simultaneous Turing and Hopf unstable modes
compete, resulting in a characteristic complex spatio-temporal pattern, where
temporal oscillations develop within each bump of a Turing pattern (Fig ~\ref{fig:turing_curves}(e)).
The existence of the intersection point of the Turing and Hopf curves and the observation of the exotic spatio-temporal patterns are robust to changes in parameters.
In passing we note that, close to onset, we could find only large amplitude patterns
of the type shown in Figs~\ref{fig:turing_curves}(c)-(e), indicating that the Turing
and Turing-Hopf bifurcations are subcritical (as confirmed below by numerical
continuation).
Interestingly, we also observe dynamic Turing patterns in region II, implying that the Hopf bifurcation does not stabilize the system, instead it creates bistability in this region, where the system supports both bulk oscillations and dynamic global patterns.

We also found other complex spatio-temporal patterns away from bifurcation onset, using direct
numerical simulation (Fig.~\ref{fig:patten_zoo}). The system supports 
time-periodic patterns containing structures within bumps  (Fig.~\ref{fig:patten_zoo}(a)). This pattern is observed as we move away from the Turing bifurcation but stay close to the Hopf curve. Structures of this type, in
which the bumps of a Turing pattern are periodically modulated in time, were observed
initially in sector IV of Fig. \ref{fig:turing_curves}(a). Spatio-temporal patterns of this form, with
modulation at the core, were
found only in parameter sets where the Turing and Hopf codimension-2
point is present, as expected. Structures within bump solutions are not seen in
standard neural field models. They are however commonly observed in spiking neuron
models \citep{Laing01a, Chow2006}, which emphasises that this next generation
neural mass model retains information about the underlying spiking model.

In region IV, we also observe an number of interesting patterns, such as the \emph{wandering} bump (Fig. \ref{fig:patten_zoo}(b)) and a form of standing wave, where both the width and the height of the bumps is periodically modulated (Fig. \ref{fig:patten_zoo}(c)). Numerical simulations indicate that the wandering bump is stable for a large area of parameter space and can coexist with a periodic wavetrain. The standing wave, however, persists for a long time but ultimately is unstable and transitions to a periodic travelling wave (not shown).

\section{Numerical bifurcation analysis}\label{sec:numeric}
In order to study patterns away from bifurcations, we employ numerical
bifurcation techniques, which allows us to compute coherent structures, determine their stability, and track their dependence on control parameters. 
Here, we employ the numerical tool kit developed by Avitabile
\cite{Avitabile2016} and employed in the context of standard neural fields
in \cite{Rankin:2014bz,Avitabile:2014wg}. This tool kit can be used to compute waves and patterns, and their stability. We refer the reader to recent reviews on
numerical bifurcation analysis for coherent structures \cite{Laing2014, Champneys2007}.
\begin{figure}
\centering
\includegraphics{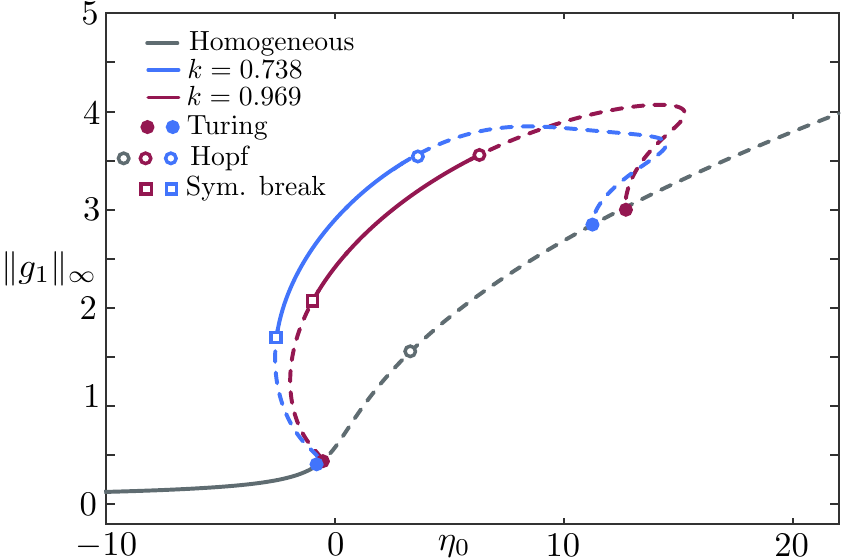}
\caption{Continuation of spatially periodic patterns. A bifurcation diagram in $\eta_0$ of two stationary patterned states, with different wavenumbers.  Each branch of patterned states are connected to the branch of homogeneous states via a subcritical Turing bifurcations and reconnect to the steady state within the region of instability. Only the rightmost and left most Turing points correspond to a change in stability of the system. The homogeneous steady state undergoes a Hopf bifurcation at $\eta_0= 3.298$ as expected from the instability analysis in \S\ref{sec:turing}. More interestingly, we also see Hopf bifurcations along the branch of periodic solutions. Solid (dashed) lines represent stable (unstable) solutions. Parameter values:
  $v_\text{syn}=15$, $\Lambda=12\pi$, other parameters as in
Fig.~\ref{fig:turing_curves}(a).}
\label{fig:bump_continuaton1}
\end{figure}
\begin{figure*}
  \centering
  \includegraphics{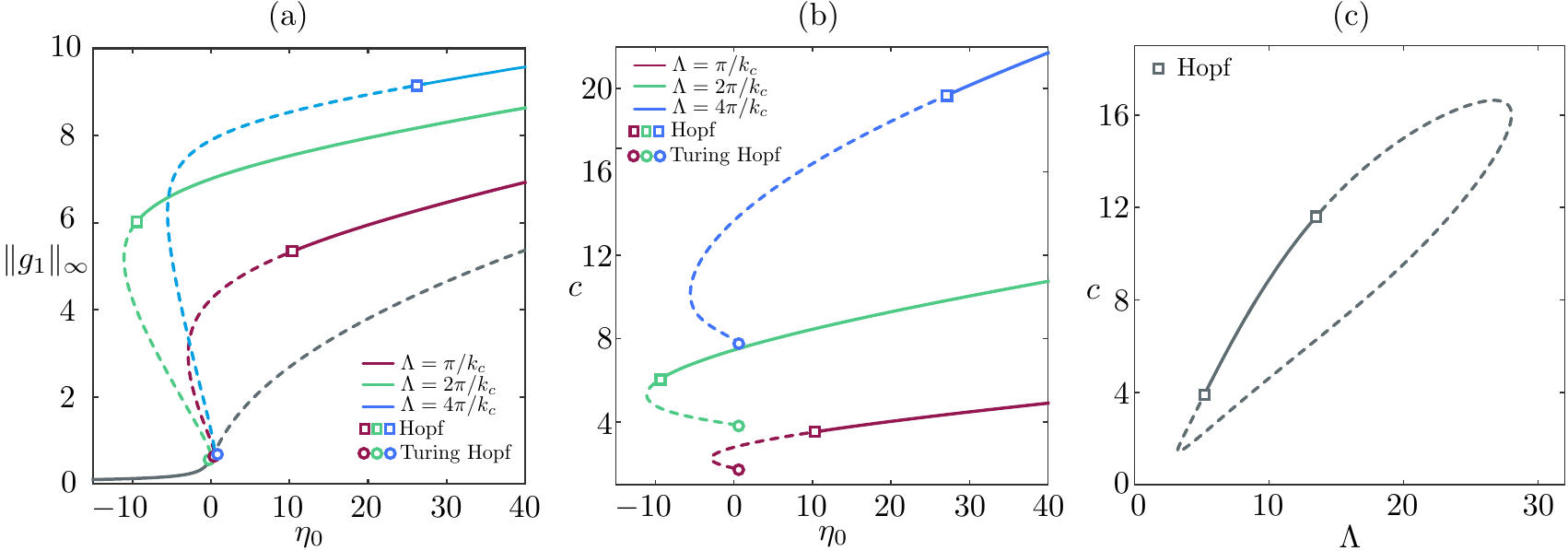}
  \caption{
  	Numerical bifurcation analysis of wavetrains. (a): Bifurcation diagram of
    wavetrains in the parameter $\eta_0$, for various values of the domain size, where $k_c=0.739$. 
	Here we show only the maximum of the 
	synaptic conductance $g_1$ as our solution measure. The green solution bifurcates off the steady state solution (grey) at a Turing-Hopf bifurcation, and the other solutions go unstable at slightly larger values of $\eta_0$. (b): the branches in (a) are plotted
    using the phase velocity $c$ as solution measure. The diagrams in (a) and (b)
    confirm that the Turing-Hopf bifurcation is subcritical. At
    the critical points, the wavetrain emerges with a non-zero phase velocity ($\omega_c/k_c$).  (c): Branches of 	
    wavetrains in the continuation parameter $\Lambda$, for
    $\eta_0=0$. This curve is effectively a
    dispersion relation, showing the wave speed as function of the spatial period. In
    this region of parameter space the branch is an isola, therefore the wavetrain
    exists only for a finite range of periods. Parameter values:
  $v_{\text{syn}}=-30$, other parameters as in Fig. \ref{fig:turing_curves}.}
  \label{fig:periodic_waves_domain_cont}
\end{figure*}
First, we rescaled space such that $x\in [-0.5, 0.5]$, to highlight the dependence on the scale of the domain size $\Lambda$. The equivalent PDE formulation~\eqref{eq:ZPDE}--\eqref{eq:GPDE} becomes
\begin{equation} 
\label{eq:WH-PDE}
\begin{aligned}
  \partial_t  z & = \big( \mathcal{F}(z) + \mathcal{G}(z,I) \big) ,\\
(\Lambda^2 - \partial_{xx})(1+\tau_1\partial_t)^2 g_1 & =  \Lambda^2\kappa_1 f(z) , \\
(\Lambda^2 - \partial_{xx}/\beta^2)(1+\tau_2\partial_t)^2  g_2 & = \Lambda^2\kappa_2 f(z).
\end{aligned}
\end{equation}

We perform numerical bifurcation analysis for heterogeneous spatially patterned steady
states and periodic travelling waves of the system above.
We construct the stationary patterns by solving for $(z,g_1,g_2)$ the boundary value problem
\begin{equation} \label{eq:WH-PDE-SS}
\begin{aligned}
  \mathcal{F}(z) + \mathcal{G}_1(z,g_1) + + \mathcal{G}_2(z,g_2) & = 0,
\\
(\Lambda^2 - \partial_{xx}) g_1 - \Lambda^2\kappa_1  f(z) & = 0, \\
(\Lambda^2 - \partial_{xx}/\beta^2) g_2 - \Lambda^2\kappa_2 f(z) & = 0, 
\end{aligned}
\end{equation}
with Neumann boundary conditions. Wavetrains are solutions
to~\eqref{eq:WH-PDE} with $z(x,t) = Z(x-ct)$, $g_1(x,t) = G_1(x-ct)$ and
$g_2(x,t) = G_2(x-ct)$ for $c \in \mathbb{R}$, where $Z(\xi)$, $G_1(\xi)$ and $G_2(\xi)$ are
$\Lambda$-periodic. To
compute wavetrains, we solve for $(Z,G_1,G_2,c)$ the boundary value problem
\begin{equation} \label{eq:WH-PDE-TW}
\begin{aligned}
  c \partial_\xi  z + \Lambda \big( \mathcal{F}(Z) + \mathcal{G}_1(Z,G_1)+ \mathcal{G}_2(Z,G_2) \big) 
  &= 0,
\\
(\Lambda^2 - \partial_{\xi\xi})(1-c\tau\partial_\xi)^2  G_1 - \Lambda^2\kappa f(Z)  &= 0, 
\\
(\Lambda^2 - \partial_{\xi\xi}/\beta^2)(1-c\tau\partial_\xi)^2  G_2 - \Lambda^2\kappa f(Z)  &= 0, 
\\
\psi(Z,G_1,G_2) &= 0,
\end{aligned}
\end{equation}
posed on $\xi \in [-1/2,1/2]$ with periodic boundary conditions. The last equation
in~\eqref{eq:WH-PDE-SS} and \eqref{eq:WH-PDE-TW} is a standard phase condition
\cite{doedel1981auto}
\begin{equation} \label{eq:phase_cond}
\begin{aligned}
  \psi(Z,G_1,G_2) & = \int_{-1/2}^{1/2} \d\xi \; \frac{\d}{\d \xi} \widetilde Z(\xi) 
		 \big( Z(\xi) - \widetilde Z(\xi) \big)\\
            &  + \int_{-1/2}^{1/2} \d\xi \; \frac{\d}{\d \xi} \widetilde G_1(\xi) 
		\big( G(\xi) - \widetilde G_1(\xi) \big)  \\
            &  + \int_{-1/2}^{1/2} \d\xi \; \frac{\d}{\d \xi} \widetilde G_2(\xi) 
		\big( G(\xi) - \widetilde G_2(\xi) \big),
\end{aligned}
\end{equation}
where $(\widetilde Z, \widetilde G_1, \widetilde G_2)$ is a reference template solution, such as one of
the solutions obtained via direct simulation. We discretised the differential
operators \eqref{eq:WH-PDE-TW} using standard differentiation matrices, which are also used to compute
linear stability of the coherent structures~\cite{Avitabile2016}.

\subsection{Turing patterns}\label{sub:bumps}
We first analyse the stationary patterns seen in \S\ref{sec:turing}, using
numerical continuation to verify the analytical results, determine the criticality of
the Turing bifurcation, and examine the behaviour of these solutions away from the
onset of the instability. We continued solutions to the boundary value
problem~\eqref{eq:WH-PDE-SS} in the parameter $\eta_0$ with $\Lambda=12\pi$, $v_{\text{syn}}=15$ and all other
parameters as in Fig.~\ref{fig:turing_curves}(a). This
corresponds to making a vertical excursion through the 2-parameter bifurcation diagram Fig.~\ref{fig:turing_curves}(a). 
Solving \eqref{eq:Turing1}--\eqref{eq:Turing2} we found two Turing bifurcations at $\eta_0=-0.648$, $k_c=0.738$ and $\eta_0=12.67$, $k_c=0.969$. Hence, we numerically continued patterns states with $k=0.738$ (blue) and $k=0.969$ (red) (Fig. \ref{fig:bump_continuaton1}). 
As expected, the homogeneous steady state bifurcates to patterns at $\eta_0=-0.648$ (blue) and $\eta_0=12.67$ (red), corresponding to the Turing bifurcations found in \S\ref{sec:turing}.
The stability of the patterned states, as well as the homogeneous state, was numerically calculated along the continuation branch. 
Along with the Turing bifurcations, the homogeneous steady state undergoes a Hopf bifurcation at $\eta_0=3.298$, which matches the value found analytically. The patterned Turing solutions were found to go unstable to a globally oscillating periodic pattern, through a Hopf bifurcation at $\eta_0=2.8380$ (blue) and $\eta_0=5.8546$ (red). The patterned solutions are also unstable (through a symmetry breaking bifurcation) to broader patterns at low values of $\eta_0$.
As anticipated, the bifurcations from the homogeneous state steady state are subcritical, hence we
have bistability between a spatial pattern and the homogeneous state and the bistability region
occurs in a wide region of parameter space. 
Further numerical continuation results
(not shown) indicate that the region of bi-stability increases as the reversal
potential $v_\text{syn}$ is increased. As $v_\text{syn}$ decreases towards $0$, the
static Turing bifurcation points collide, and the patterned states cease to exist, as predicted from the Turing analysis in \S\ref{sec:turing}. We found the scenario presented above to be robust to changes in the other parameters.

\subsection{Wavetrains}\label{subsub:moiving_bumps}
We now shift our focus to wavetrain solutions originating at a Turing-Hopf
bifurcation of the homogeneous steady state. We recall that we find these states as
solutions to~\eqref{eq:WH-PDE-TW} with periodic boundary conditions, hence $\Lambda$
corresponds to the spatial period of the wavetrain profile. In addition, the phase
velocity $c$ of a wavetrain is accessible from the boundary-value problem solution.


As in \S\ref{sub:bumps}, the spatial frequency $k_c$ at bifurcation was found by solving 
\eqref{eq:dynTuring1}--\eqref{eq:dynTuring3}, and we continued patterns with this wave number by setting $\Lambda=2\pi/k_c$, where $k_c=0.739$ (Fig.
\ref{fig:periodic_waves_domain_cont}(a)--(b) green curve). We also continued spatial patterns with $k=2 k_c$ (red curve) and $k=k_c/2$ (blue curve), this was achieved by posing the system on domains $\Lambda=\pi/k_c$ and $\Lambda=4\pi/k_c$, respectively.
The wavetrains bifurcate from the homogeneous 
steady state with a non-zero phase velocity, at different values of $\eta_0$ (Fig.~\ref{fig:periodic_waves_domain_cont}(b)). The
value for which periodic waves emerge for $\Lambda=2\pi/k_c$ 
corresponds to the Turing-Hopf bifurcation (green dot) found in \S\ref{sec:turing}, and the speed is equal to $\omega_c/k_c$.
The wavetrains are unstable for low values of $\eta_0$. For the smaller domain size they are unstable to the stable steady state, whereas for the larger domain size they transition to finer patterns, with a smaller spatial wavelength $k$.

To fully explore the relationship between the wave speed and the spatial period, we
fix the value of $\eta_0=0$ and use $\Lambda$ as a continuation parameter. This 
opens up the possibility to trace branches of solutions in the $(c,\Lambda)$-plane, 
and hence, approximate the dispersion curve of the waves (Fig.
\ref{fig:periodic_waves_domain_cont}(c)). We found that wavetrains occur in isolas,
and therefore only exist for a finite range of spatial periods. For $5.2<\Lambda<13.1$, a fast (stable) wavetrain coexist with slower (unstable) one. Whereas, when $
3.2<\Lambda<5.2 $ and $13.1<\Lambda<28.0 $ the two waves are unstable. For all other values of $\Lambda$ we do not see wavetrain solutions. If $\eta_0$ is increased, the dispersion curve is no longer an isola, but rather a 
monotonically increasing function. Thus, in this regime, the system supports periodic travelling waves 
for all values of the spatial period, above a threshold value.

\subsection{Fronts}\label{sub:waves}
\begin{figure*}
  \centering
  \includegraphics{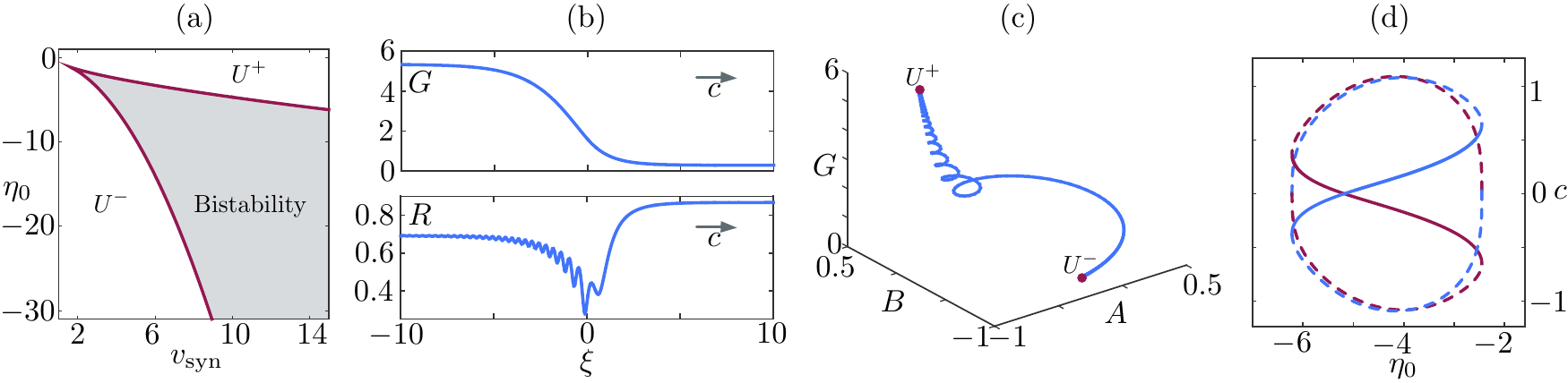}
  \caption{Numerical continuation results for travelling fronts. (a): two-parameter
  bifurcation diagram of stationary homogeneous states of~\eqref{eq:EK-PDE}
  (equilibria of~\eqref{eq:spatialDynamic}) in the $(\eta_0,v_{\text{syn}})$-plane. We plot
  the locus of saddle-node bifurcations of the steady state, colliding at a cusp
  bifurcation. To the left of the cusp we identify a low-activity state, $U^-$, and a
  high-activity one, $U^+$, which coexist and are stable in the shaded area.
  Parameter values: $\Delta=0.5$, $\kappa=5$, $\tau=1$. (b) Travelling front
  profile, computed by solving~\eqref{eq:spatialDynamic} on a truncated domain
  $[-30,30]$, for $\eta_0=-3$, $v_{\text{syn}}=4$. We show the profiles in the synaptic
  conductance variable $G(\xi)$ and the synchrony variable $R(\xi) = |Z(\xi)|$. (c)
  The travelling front in (b) is a heteroclinic orbit connecting the equilibria $U^+$
  and $U^-$ of the spatial-dynamical system~\eqref{eq:spatialDynamic}. We show a
  projection of the (approximate) heteroclinic orbit in the $(A, B,
  G)$-space. (d) Numerical continuation of the travelling front $U(\xi)$ found in (b) (blue) and its symmetric counter $U(-\xi)$ (red), using
  $\eta_0$ as continuation parameter and $c$ as solution measure. The fronts live on an
  isolated branch and destabilise at saddle-node bifurcations. The bifurcation curve is
  symmetric, with respect to the axis $c=0$. Solutions with $c>0$ and $c<0$ are related
  via the transformation $U(\xi)\mapsto U(-\xi)$. }
  \label{fig:front_continuation}
\end{figure*}

Standard neural field models are known to support travelling fronts (for exponentially decaying kernels), which are
travelling waves whose profile connects a uniform high-activity state to another low-activity state \cite{Ermentrout1998, Bressloff2012a}. It is therefore natural to search for these coherent
structures in our new  neural field model~\eqref{eq:ZEvol}--\eqref{eq:GEvol}. As in any
other nonlocal neural field, the existence of the high- and low-activity states
depends on the choice of the synaptic kernel. 
As inhibition is not necessary for the generation of localised patterns, such as travelling fronts, in this section we consider only a single synaptic conductance, $m=1$ and suppress the $m$ label, leading to the equivalent PDE formulation
\begin{equation} \label{eq:EK-PDE}
\begin{aligned}
  \partial_t  z & = \mathcal{F}(z) + \mathcal{G}(z,g),
\\
(1 - \partial_x^2)(1+\tau\partial_t)^2 g & = \kappa f(z) ,
\end{aligned}
\end{equation}
which we pose on $\xi \in \mathbb{R}$. 
We now set $z(x,t) = Z(x-ct)$, $g(x,t) =
G(x-ct)$, and we seek travelling fronts as bounded solutions $U(\xi) = (U_1(\xi), \ldots, U_6(\xi))$ to the
boundary-value problem
\begin{equation}\label{eq:spatialDynamic}
  \partial_\xi U = \mathcal{N}(U), \quad \xi \in \mathbb{R}, \qquad \lim_{\xi \to \pm
  \infty} U(\xi) = U^{\mp},
\end{equation}
where $\mathcal{N} \colon \mathbb{R}^6 \to \mathbb{R}^6$ is the real-valued nonlinear
function
\[
  \mathcal{N}(U)=
  \begin{pmatrix}
    -\real [ \mathcal{F}( U_1 + i U_2) + \mathcal{G}( U_1 +
	i U_2, U_3) ]/c \\
    -\imag [ \mathcal{F}( U_1 + i U_2) +  \mathcal{G}( U_1 +
	i U_2, U_3) ]/c \\
    (U_3 - U_4)/(\tau c) \\
    U_5 \\ 
    U_6 \\ 
    U_5 + \big[ U_6 + \kappa f(U_1 + i U_2) - U_4\big]/(\tau c)
  \end{pmatrix}.
\]
In this spatial-dynamical system formulation of the problem, the first three
components of $U$ have a direct interpretation in terms of the state variables
$(z,g)$ of \eqref{eq:EK-PDE},
\[
  U_1 = A \equiv \real Z, \qquad U_2 = B \equiv \imag Z, \qquad U_3 = G,
\]
whereas $U_4$, $U_5$ and $U_6$ are auxiliary variables, necessary to cast the problem
as a system of first-order differential equations in $\xi$. The equilibria 
$
  U^\pm = (U_1^\pm,U_2^\pm,U_3^\pm,U_3^\pm,0,0)
$
of the spatial-dynamical system~\eqref{eq:spatialDynamic} correspond to high- and
low-activity homogeneous steady states of~\eqref{eq:EK-PDE} and are completely
determined by solving for $(U_1,U_2,U_3)$ the algebraic problem
\begin{equation}\label{eq:clampedEq}
  \begin{aligned}
    \real [ \mathcal{F}( U_1 + i U_2) + \mathcal{G}( U_1 +
	i U_2, U_3) ] & = 0,\\
    \imag [ \mathcal{F}( U_1 + i U_2) +  \mathcal{G}( U_1 +
	i U_2, U_3) ] & = 0,\\
    \kappa f(U_1 + i U_2) - U_3 & = 0.
  \end{aligned}
\end{equation}
We define $U^+$ as the high activity state, which displays high synaptic conductance and $U^-$ as the low activity state, displaying lower synaptic conductance. Note that we also seek the symmetric counterpart solution where $\lim_{\xi \to \pm
  \infty} U(\xi) = U^{\pm}$. 
We have continued solutions to~\eqref{eq:clampedEq} in $\eta_0$ and $v_\text{syn}$
using XPPAUT~\cite{Ermentrout02} (Fig. \ref{fig:front_continuation}(a)). 
The system has three fixed points in the region enclosed by the saddle-node curves, two
of which are stable ($U^\pm$). Therefore, we look for travelling waves in this shaded
region of parameter space. We note that, as $\tau$ is decreased, a Hopf bifurcation of the $U^-$ state arises, opening up the possibility of creating
heteroclinic connections to periodic orbits, which we will not consider further here.

The state at $\xi\to-\infty$ ($\xi \to \infty$) displays a
high-conductance (low-conductance) $G$, hence a high (low) firing, and it is
therefore referred to as the high (low) activity state (Fig. \ref{fig:front_continuation}(b)). We computed travelling waves using the
routines from~\cite{Avitabile2016}, which allows us to study the spectral stability of the
waves. Computations are performed on a large truncated domain $\Lambda = 60$,
and plotted on $[-10,10]$ for convenience. Interestingly, we see ripples in the wake
of the front, which indicates that the solution connects a node to a focus. The solution
in the phase space $(A,B,G)$ illustrates that the high-activity state ($U^-$) is a focus, and 
the low activity one ($U^+$) a node.

We continued the travelling wave shown in Fig.~\ref{fig:front_continuation}(b) (blue), and its symmetric counterpart (red), in the
mean background drive $\eta_0$, using $c$ as a solution measure (Fig.~\ref{fig:front_continuation}(d)). Fronts live on an isolated branch and destabilise
at saddle-node bifurcations. The bifurcation diagram is symmetric with
respect to the axis $c=0$, as solutions on the branch with $c>0$ and
$c<0$ are related via the transformation $U(\xi) \mapsto U(-\xi)$. Up to $6$
coexisting waves exist in a wide region of $\eta_0$ parameter space, albeit only two
of them are stable. In addition, we found stable waves in which the high activity state $U^+$ is moving across the tissue, invading the low activity state $U^-$, and vice versa. The inversion of velocity occurs where the blue and red
stable branches overlap. 

\begin{figure*}
\centering
\includegraphics{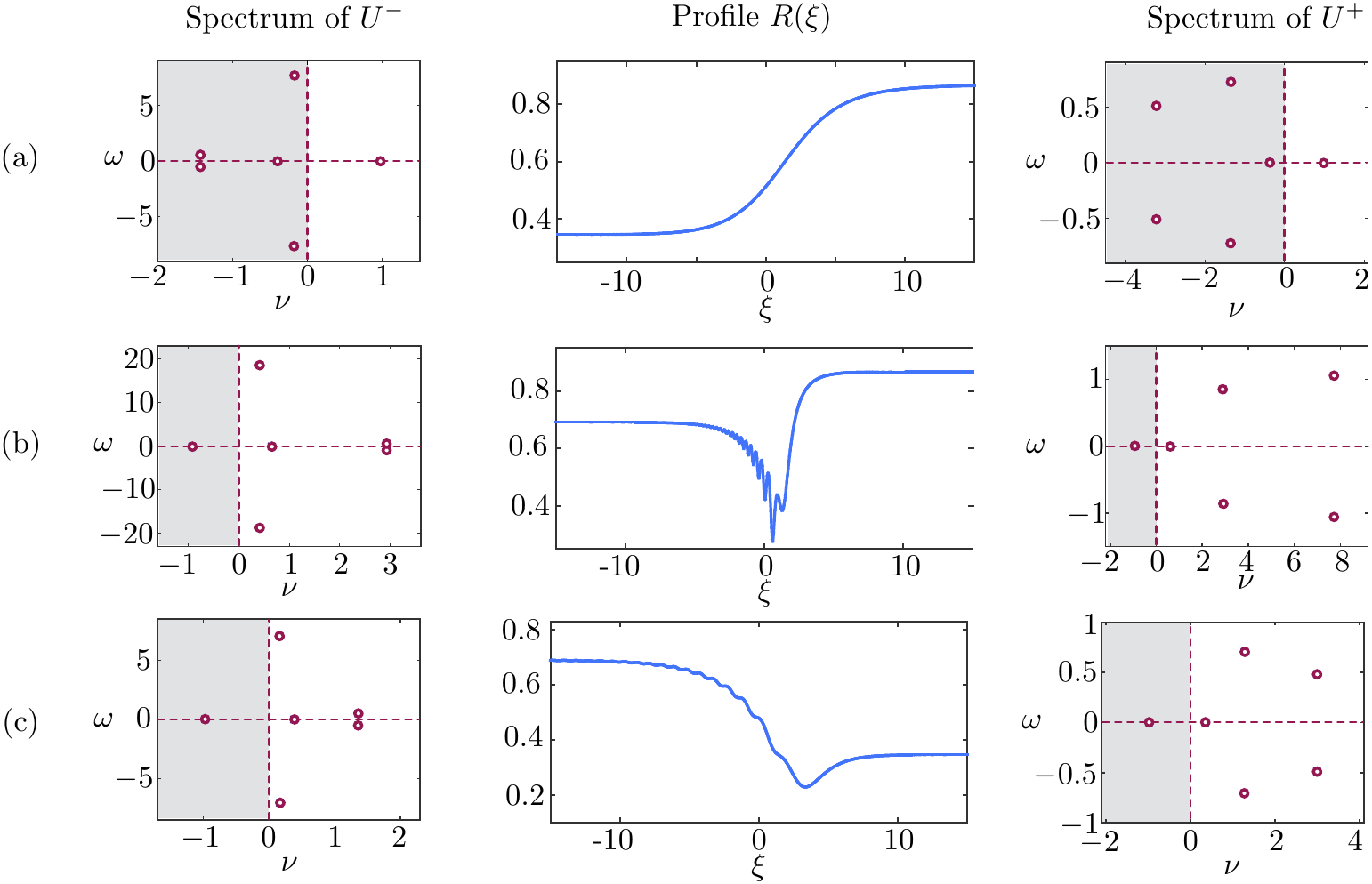}
\caption{Spectrum of the high- and low-activity homogeneous states connected by
  selected travelling wave profiles, in Fig. \ref{fig:front_continuation}. (a)
  $c=-0.8769$, (b) $c=0.3594$, (c) $c=0.9470$.  Parameter values: $\eta_0=-3$,
  $v_\text{syn}=4$, $\Delta=0.5$, $\kappa=5$, $\tau=1$.}
\label{fig:wave_connections}
\end{figure*}

As the isola of travelling fronts is traced, these solutions gain or lose
oscillations in the wake of the wave. To investigate further this aspect, we monitor
the spectrum of $U^\pm$ (as equilibria of the spatial-dynamical
system~\eqref{eq:spatialDynamic}) as we move along the branch. Taking a vertical excursion in Fig.~\ref{fig:front_continuation}(d), we find three independent solutions, one of which is stable and two of which unstable.  We examine the stability of the fixed points $U^\pm$ in the travelling wave frame for each of these solutions. As expected,
oscillations in the wake of the wave are absent where the unstable spectrum of
$U^-$ is purely real (Fig.~\ref{fig:wave_connections}(a)), and they develop when a complex-conjugate pair of eigenvalues
crosses the imaginary axis (Fig.~\ref{fig:wave_connections}(b)). We also find that the amplitude of the oscillations
is small when the real part of the complex eigenpair is small (Fig.~\ref{fig:wave_connections}(c)).

\section{\label{sec:discussion}Discussion}
We have presented the derivation of an atypical neural field model from a network of
spatially distributed $\theta$-neurons. In the reduced model, which we dub a
\emph{next generation neural field model}, within population synchrony drives the
population firing rate. The new model supports a range of patterns, such as bumps,
waves and breathers. Noteworthy is the state characterised by structures within
bumps, as these states are not seen in standard neural mass models. These structures
instead typify patterns seen in networks of spiking neurons, which signifies that by
maintaining the notion of within population synchrony this neural field model can retain
information about the underlying spiking network. Exotic states, with within bump oscillations, were found in
the region where the Hopf and Turing bifurcations collided.

A Turing instability analysis provided us with an understanding of how the system
behaved close to bifurcation points, allowing us to determine when the system
transitioned from the homogeneous steady state.
However, unlike the Amari model we cannot use the Heaviside approximation to make
further analytical progress since the firing rate is now a fixed real valued function
of the Kuramoto order parameter. As such, we have moved to numerical continuation
techniques to analyse the behaviour of the system away from these bifurcation points.
Numerical techniques were also used to examine the existence and stability of
travelling fronts.

Previous work \cite{Byrne2017} illustrated that the point version of this model with an external time-dependent drive (without spatial extent)
could support $\beta$-rebound, an event-related modulation of the beta rhythm, as
seen in MEG. An interesting extension of this work would be to include an additional
drive in the neural field model to examine how the inclusion of space affects $\beta$
band modulations, and perhaps explain why $\beta$-rebound is seen in both the
contralateral and the ipsilateral hemispheres during movement. More generally the
model parameters can be altered so that the population oscillates at other
frequencies, and hence, used to explain other event-related
desynchronisation/synchronisation phenomena in the brain.

\begin{figure*}
  \centering
  \includegraphics{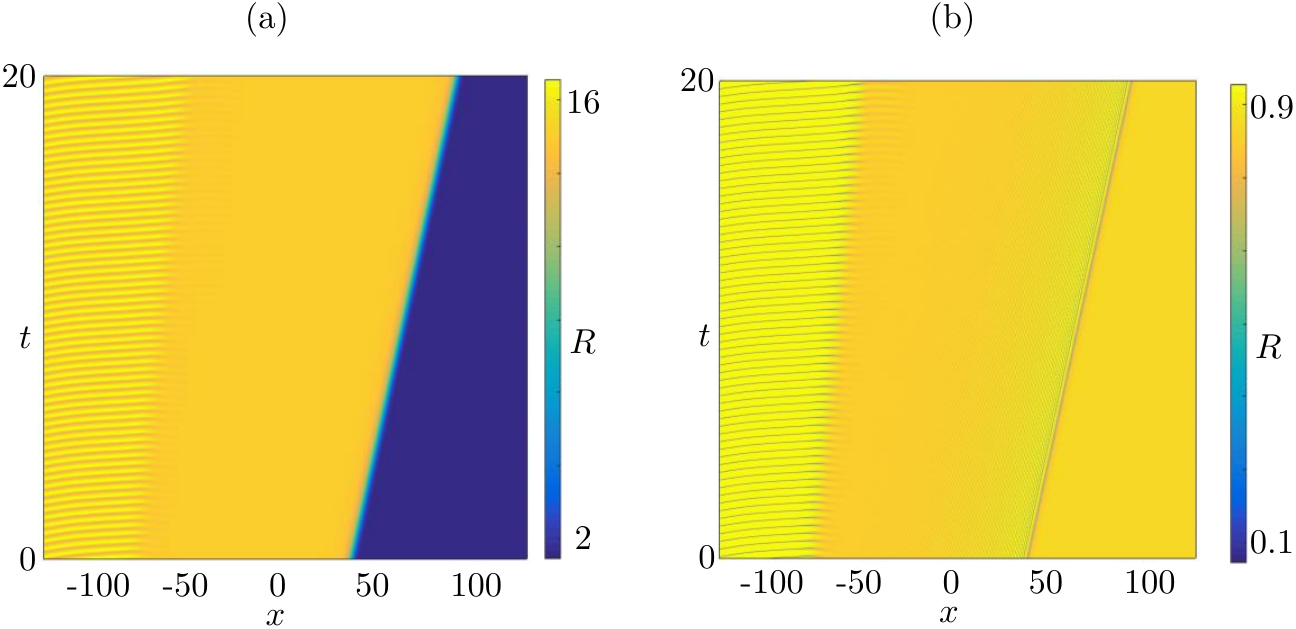}
  \caption{Travelling wave connecting a periodic orbit to a node: Surface plots
    showing the evolution of (a) the synaptic conductance $g$ and (b) the synchrony
    $R$ for a front which connects an oscillatory state to a fixed point state.
    Simulations for the system defined by \eqref{eq:EK-PDE}, with $\eta_0=-5$,
  $v_\text{syn}=10$, $\Delta=0.5$, $\kappa=5$, $\tau=0.2$
    }
  \label{fig:front_orbits}
\end{figure*}

In \S\ref{sec:numeric}, we pointed to the existence of a front which connects periodic
orbits to nodes/focuses, for the model with an exponential coupling kernel. We 
observed such fronts by decreasing the synaptic time constant $\tau$ (Fig.
\ref{fig:front_orbits}). The numerical machinery used here doesn't allow for the 
continuation of these solutions, known as
defects. The analysis of defects is still an open problem. Close examination of Fig.
\ref{fig:front_orbits} reveals that there are two fronts in the connection between the
node and the limit cycle, which appear to be moving at different speeds. Even more
interesting, would be the analysis of fronts which connect two periodic orbits of
different amplitudes. The spreading of such a wave across the cortex could be viewed
as the spreading of an epileptic seizure.
Another numerical challenge would be to continue the exotic patterned states seen in
\S\ref{sec:turing} when the Turing and Hopf bifurcations collide. These
patterns have both a spatial and a temporal period, which would require extending the
numerical machineries \citep{Avitabile2016} to continue both a
spatial and a temporal pattern. This has been achieved in \citep{Avitabile2008} for the Brusselator model.

A natural extension to the work presented in both \S\ref{sec:turing} and
\S\ref{sec:numeric} would be to include a second spatial
dimension. It is more natural to view the cortex as a two dimensional sheet and
examine the propagation of waves across it. We would expect that the 2D system
supports the two dimensional versions of the patterned states presented here, but
also potentially some more exotic states. Extending both the Turing analysis and the
numerical machinery to include a second spatial extension is worthy of further exploration.

\section{Acknowledgements}
DA was supported by the Engineering and Physical Sciences Research Council under
grant EP/P510993/1. SC was supported by the European Commission through the FP7 Marie
Curie Initial Training Network 289146, NETT: Neural Engineering Transformative
Technologies.

\appendix

\section{Jacobian}
To calculate the Jacobian we first write the system \eqref{eq:ZEvol}--\eqref{eq:GEvol} in its full six dimensional form,
\begin{align*}
\PD{a}{t} &=b(a-1)-b(a+1)(\eta_0+v_{\text{syn}}^1 g_1+v_{\text{syn}}^2 g_2)\\
&\phantom{.}\hspace{1.5em}-(a+1)\Delta-\frac{1}{2}(a^2-b^2-1)(\Delta+g_1+g_2),  \\
\PD{b}{t} &=-\frac{1}{2}((a-1)^2-b^2)-b\Delta-ab(\Delta+g_1+g_2) \\
&\phantom{.}\hspace{1.5em}+\frac{1}{2}((a+1)^2-b^2)(\eta_0+v_{\text{syn}}^1 g_1+v_{\text{syn}}^2 g_2) ,
\end{align*}
\begin{align*}
\PD{g_1}{t} &= \frac{1}{\tau_1}(-g_1 + K_1) ,\\
\PD{K_1}{t} &= \frac{1}{\tau_1}(-K_1 + \kappa_1 w_m\otimes f(a+ib)), \\
\PD{g_2}{t} &= \frac{1}{\tau_2}(-g_2 + K_2), \\
\PD{K_2}{t} &= \frac{1}{\tau_2}(-K_2 + \kappa_2 w_m\otimes f(a+ib)).
\end{align*}
The Jacobian of the system can be written as follows:
\label{app:jacobian}
\begin{equation*}
\mathcal{J}(k) = \left(\begin{matrix}
\mathcal{J}_{11} & \mathcal{J}_{12} \\
\mathcal{J}_{21}(k) & \mathcal{J}_{22}
\end{matrix}\right),
\label{jacobian}
\end{equation*}
where,
\begin{equation*}
\mathcal{J}_{11} = \left(\begin{matrix}
\PD{}{a}\PD{a}{t} \hspace{0.5em}& \PD{}{b}\PD{a}{t}  \vspace{1em}\\
\PD{}{a}\PD{b}{t}  \hspace{0.5em}& \PD{}{b}\PD{b}{t} 
\end{matrix}\right),
\end{equation*}
\begin{equation*}
\mathcal{J}_{12} = \left(\begin{matrix}
 0 \hspace{0.5em}& \PD{}{g_1}\PD{a}{t} \hspace{0.5em}& 0\hspace{0.5em} &\PD{}{g_2}\PD{a}{t} \vspace{1em}\\
 0 \hspace{0.5em}& \PD{}{g_1}\PD{b}{t} \hspace{0.5em}& 0 \hspace{0.5em}& \PD{}{g_2}\PD{b}{t} \vspace{1em}
\end{matrix}\right),
\end{equation*}
\begin{equation*}
\mathcal{J}_{21}(k)  = \left(\begin{matrix}
\tau_1^{-1}\kappa_1 \widehat{w}_1\PD{f}{a}\hspace{0.5em} \phantom{.}& \tau_1^{-1}\kappa_1 \widehat{w}_1\PD{f}{b} \vspace{1em}\\
0 & 0 \vspace{0.5em}\\
\tau_2^{-1}\kappa_2 \widehat{w}_2\PD{f}{a}  \hspace{0.5em} \phantom{.}& \tau_2^{-1}\kappa_2 \widehat{w}_2\PD{f}{b}\vspace{1em}\\
0 & 0
\end{matrix}\right),
\end{equation*}
\begin{equation*}
\mathcal{J}_{22} = \left(\begin{matrix}
 -\tau_1^{-1} & 0 & 0 & 0 \vspace{0.5em}\\
\tau_1^{-1} & -\tau_1^{-1} & 0 & 0 \vspace{0.5em}\\
 0 & 0 & -\tau_2^{-1} & 0 \vspace{0.5em}\\
 0 & 0 & \tau_2^{-1} & -\tau_2^{-1}
\end{matrix}\right).
\end{equation*}
The variables, $a$, $b$, $g_1$ and $g_2$ are evaluated at the steady state,
and hence, depend upon the control parameters, $\widehat{w}_i$ depend on $k$ and $\beta$ and $f$ is given by \eqref{eq:firing_rate}.

The derivatives are computed as follows:
\begin{align*}
&\PD{}{a}\left(\PD{a}{t}\right) = \phantom{-}\PD{}{b}\left(\PD{b}{t}\right) ,\\
&= b-(a+1)\Delta-b(\eta_0+v_{\text{syn}}^1 g_1+v_{\text{syn}}^2 g_2)-a(g_1+g_2), \\
&\PD{}{b}\left(\PD{a}{t}\right) = -\PD{}{a}\left(\PD{b}{t}\right), \\
&= (a-1)-(a+1)(\eta_0+v_{\text{syn}}^1 g_1+v_{\text{syn}}^2 g_2)+b(\Delta+g_1+g_2),\\
&\PD{}{g_i}\left(\PD{a}{t}\right)  = b(a+1)v_{\text{syn}}^i - (a^2-b^2-1), \\
&\PD{}{g_i}\left(\PD{b}{t}\right)  = \frac{1}{2}((a+1)^2-b^2)v_{\text{syn}}^i -ab, \\
&\PD{f}{a} = -\frac{2}{\pi}\frac{(a+1)^2-b^2}{((a+1)^2-b^2)^2},\\
&\PD{f}{b} = -\frac{4}{\pi}\frac{b(a+1)}{((a+1)^2-b^2)^2}.
\end{align*}

\section{Turing coefficients}
\label{app:coeff}
The coefficients of the characteristic equation \eqref{eq:characteristic_formula} in \S\ref{sec:turing} are calculated as follows
\begin{align*}
p_0 &= \left[\mathcal{A}^2+\mathcal{B}^2\right]/(\tau_1\tau_2)^2, \\
p_1 &=   2\left[(\tau_1+\tau_2)\left(\mathcal{A}^2+\mathcal{B}^2\right) - \mathcal{A}\right]/(\tau_1\tau_2)^2, \\
\end{align*}
\begin{align*}
p_2 &= \left[(\tau_1^2 +\tau_2^2+ 4\tau_1\tau_2)\left(\mathcal{A}^2+\mathcal{B}^2\right)+1 \right.\\
&\phantom{.}\hspace{8em}\left.-4(\tau_1+\tau_2)\mathcal{A}\right]/(\tau_1\tau_2)^2,\\
p_3 &= 2\left[(\tau_1\tau_2^2+\tau_1^2\tau_2)\left(\mathcal{A}^2+\mathcal{B}^2\right) + (\tau_1+\tau_2 -4\mathcal{A})\right. \\
&\phantom{.}\hspace{8em} \left.- (\tau_1^2 + \tau_2^2)\mathcal{A}\right]/(\tau_1\tau_2)^2,\\
p_4 &= \tau_1^{-2} + \tau_2^{-2} +\mathcal{A}^2+\mathcal{B}^2 + 4\left(1 -(\tau_1+\tau_2)\mathcal{A} \right)/(\tau_1\tau_2), \\
p_5 &= 2 \left(\tau_1^{-1} + \tau_2^{-1} -\mathcal{A}\right), \\
q_0^i &= \kappa_i\mathcal{C}_i/(\tau_i\tau_j)^2,\\
q_1^i &= \left[\kappa_i f(a,b) +2\tau_j\kappa_i\mathcal{C}_i\right]/(\tau_i\tau_j)^2, \\
q_2^i &= \left[2\kappa_i f(a,b) +  \tau_j\kappa_i\mathcal{C}_i\right]/(\tau_i^2\tau_j), \\
q_3^1 &= \kappa_i f(a,b)/\tau_i^2,
\end{align*}
where $i,j \in\{1,2\}$ and
\begin{align*}
\mathcal{A} &\equiv \PD{}{a}\left(\PD{a}{t}\right) = \phantom{-}\PD{}{b}\left(\PD{b}{t}\right) \\
\mathcal{B} &\equiv \PD{}{b}\left(\PD{a}{t}\right) = -\PD{}{a}\left(\PD{b}{t}\right) \\
\mathcal{C}_i &= -\mathcal{A}f(a,b)+\frac{\mathcal{B}}{\pi}\left(v_{\text{syn}}^i - \frac{2b}{(1+a)^2+b^2}\right).
\end{align*}

\bibliography{references}

\end{document}